\documentclass[aps,pra,twocolumn,groupedaddress,showpacs]{revtex4}

\usepackage{graphicx}

\begin{document}

%======================== CROSS-REFERENCING ==================
\newcommand{\eq}[1]{Eq.~(\ref{eq.#1})} % Ref. to equation
\newcommand{\eqbare}[1]{(\ref{eq.#1})} % Ref. to equation number only

\newcommand{\fig}[1]{Fig.~\ref{fig.#1}}
\newcommand{\figbare}[1]{\ref{fig.#1}}

\newcommand{\eqlabel}[1]{\label{eq.#1}}
\newcommand{\figlabel}[1]{\label{fig.#1}}

%======================== common expressions ==================
\newcommand{\expect}[1]{ \left< #1 \right> }
\newcommand{\floor}[1]{ {\lfloor #1 \rfloor } }
\newcommand{\Hm}{ H^{\rm (0)} }
\newcommand{\Hc}{ H^{\rm (c)} }
\newcommand{\Psoln}{ P_{\rm soln} }
\newcommand{\Nclauses}{ M } % total number of possible clauses to select from

\newcommand{\AtMost}[1]{ O(#1) }
\newcommand{\AtLeast}[1]{ \Omega(#1) }

\newcommand{\comment}[1]{#1} % for notes not included in published version

%\preprint{}

\title{Adiabatic Quantum Computing for Random Satisfiability Problems\footnote{appeared in {\it Phys Rev A} {\bf 67}, 022314 (2003)}}
\author{Tad Hogg}
%\email[email:]{tad_hogg@hpl.hp.com}
%\homepage[]{http://www.hpl.hp.com/shl/people/tad}
%\thanks{}
\affiliation{HP Labs%, MS 1139\\1501 Page Mill Road
    \\Palo Alto, CA 94304}

%\date{}

\begin{abstract}
The discrete formulation of adiabatic quantum computing is compared
with other search methods, classical and quantum, for random
satisfiability (SAT) problems. With the number of steps growing only
as the cube of the number of variables, the adiabatic method gives
solution probabilities close to 1 for problem sizes feasible to
evaluate via simulation on current computers. However, for these sizes
the minimum energy gaps of most instances are fairly large, so
the good performance scaling seen for small problems may not reflect
asymptotic behavior where costs are dominated by tiny gaps. Moreover,
the resulting search costs are much higher than for other
methods. Variants of the quantum algorithm that do not match the
adiabatic limit give lower costs, on average, and slower growth than
the conventional GSAT heuristic method.
\end{abstract}

% 03.67.Lx Quantum Computation
%\keywords{}

\maketitle

% If in two-column mode, this environment will change to single-column
% format so that long equations can be displayed. Use
%\begin{widetext}
%\end{widetext}

\section{INTRODUCTION}

Quantum computers~\cite{deutsch85,divincenzo95,feynman96,steane98} can
rapidly evaluate all search states of nondeterministic polynomial (NP)
problems~\cite{garey79}, but appear unlikely to give short worst-case
solution times~\cite{bennett94}. Of more practical interest is whether
their average performance improves on conventional
heuristics. 

Adiabatic quantum computing, using a slowly changing time-dependent
Hamiltonian, appears to give polynomial average cost growth for some
NP combinatorial search problems~\cite{farhi01}. These observations,
while encouraging, are limited to small problems for which other
methods, both conventional and quantum, can have even lower
costs. Furthermore, although adiabatic methods apparently show
exponential cost scaling for set partitioning~\cite{smelyanskiy02} and
finding the ground state of spin glasses~\cite{santoro02}, the typical
performance of adiabatic quantum computing for large NP search
problems remains an open question. Thus it is of interest to compare
the adiabatic method with other techniques for NP problems having a
well-studied class of hard instances.

This paper provides such a comparison for $k$-satisfiability
($k$-SAT), consisting of $n$ Boolean variables and $m$ clauses. A
clause is a logical OR of $k$ variables, each of which may be
negated. A solution is an assignment, i.e., a value, true or false,
for each variable, satisfying all the clauses.
%An assignment is said to conflict with any clause it doesn't satisfy.
An example 2-SAT instance with 3 variables and 2 clauses
is $v_1$ OR (NOT $v_2$) and $v_2$ OR $v_3$, which has 4 solutions,
e.g., $v_1=v_2={\rm false}$ and $v_3={\rm true}$.
%For assignments $r$ and $s$, which can be viewed as bit-vectors of
%length $n$, let $d(r,s)$ be the Hamming distance between them, i.e.,
%the number of variables they assign different values.
For a given instance, let the cost $c(s)$ of an assignment $s$ be
the number of clauses it does not satisfy.
%Solutions are states for which $c(s)=0$.

For $k\ge 3$, $k$-SAT is NP-complete~\cite{garey79}, i.e., among
the most difficult NP problems in the worst case. For average
behavior we use the random $k$-SAT ensemble, in which the $m$ clauses
are selected uniformly at random. I.e., for each clause, a set of
$k$ variables is selected randomly, and each selected variable is
negated with probability $1/2$.
\comment{Thus clauses are selected uniformly from among the $\Nclauses = {n \choose k} 2^k$ possible clauses.
We focus on the decision problem, i.e., finding a solution, rather than the related optimization problem, i.e., finding a minimum cost state. This allows direct comparison with prior empirical studies of heuristic methods for SAT.
}
The algorithms we consider are probabilistic, so cannot definitively
determine no solution exists. Thus we use soluble instances: after
random generation, we solve the instances with an exhaustive
conventional method and only retain those with a solution.  This
ensemble has a high concentration of hard instances near a phase
transition in search
difficulty~\cite{cheeseman91,kirkpatrick94,hogg96d,monasson99}.
For 3-SAT, we generate instances near this transition by using $\mu \equiv m/n
= 4.25$, though for those $n$ not divisible by 4, half
the samples had $m=\floor{4.25 n}$ and half had $m$ larger by 1.

The remainder of this paper describes several quantum search
algorithms in the context of satisfiability problems, and then
compares their behavior.

\section{ALGORITHMS}

The adiabatic technique~\cite{farhi01} is based on two Hamiltonians
$\Hm$ and $\Hc$. The first is selected to have a known ground state,
while the ground states of $\Hc$ correspond to the solutions of the
problem instance to be solved. The algorithm continuously evolves the
state of the quantum computer using $H(f)=(1-f)\Hm + f \Hc$ with $f$
ranging from 0 to 1.  Under suitable conditions, i.e., with a nonzero
gap between relevant eigenvalues of $H(f)$, the adiabatic theorem
guarantees that, with sufficiently slow changes in $f$, the evolution
maps the ground state of $\Hm$ into a ground state of $\Hc$, so a
subsequent measurement gives a solution. The choices of $\Hm$, $\Hc$
and how $f$ varies as a function of time are somewhat arbitrary.

In matrix form, one Hamiltonian with minimal-cost assignments as
ground states is $\Hc_{r,s} = c(s) \delta_{r,s}$, for assignments $r$
and $s$, where $\delta_{r,s}$ is 1 if $r=s$ and 0 otherwise. This
Hamiltonian introduces a phase factor in the amplitude of assignment
$s$ depending on its associated cost $c(s)$.

For $\Hm$, we introduce a nonnegative weight $w_i$ for variable $i$,
let $\omega \equiv \sum_{i=1}^n w_i$ and take
\begin{equation}\eqlabel{Hmix}
\Hm_{r,s} = \cases{
    \omega/2    & \small  if $r=s$ \cr
    -w_i/2      & \small  if $r$ and $s$ differ only for variable $i$\cr
    0       & \small  otherwise \cr
}
\end{equation}
This Hamiltonian can be implemented with elementary quantum gates by
use of the Walsh-Hadamard transform $W$, with elements
$W_{r,s}=2^{-n/2}(-1)^{r \cdot s}$ (treating the states $r$ and $s$ as
vectors of bits so their dot product counts the number of variables
assigned the value 1 in both states). Specifically, $\Hm = W D W$
where $D$ is a diagonal matrix with the value for state $r$ given by the
weighted sum of the bits: $\sum_{i=1}^n w_i r_i$ with $r_i$
representing the value of the $i^{th}$ bit of $r$. In particular, if
all the weights equal 1, $D_{r,r}$ just counts the number of bits
equal to 1.

The adiabatic method is a continuous process. To compare with
other algorithms, we use the algorithmically equivalent discrete
formulation~\cite{farhi00,vandam01} acting on the amplitude vector
initially in the ground state of $\Hm$, i.e.,
$\psi^{(0)}_s=2^{-n/2}$. This formulation consists of $j$
steps and a parameter $\Delta$. Step $h$ is a matrix multiplication:
\begin{equation}\eqlabel{step}
\psi^{(h)} = e^{-i \tau(f) \Hm \Delta} \; e^{-i \rho(f) \Hc \Delta} \;  \psi^{(h-1)}
\end{equation}
with the mixing phase function $\tau(f)=1-f$, cost phase function
$\rho(f)=f$ and taking $\hbar=1$. After these steps, the probability
to find a solution is $\Psoln = \sum_s |\psi^{(j)}|^2$, with the sum
over all solutions $s$.

As a simple choice for the evolution, we take $f$ to vary linearly
from 0 to 1. We exclude the steps with $f=0$ and $1$ since they have
no effect on $\Psoln$. Specifically, we take $f=h/(j+1)$ for step $h$,
ranging from 1 to $j$.

The expected number of steps required to find a solution is
$C=j/\Psoln$, providing a commonly used proxy for the computational
cost of discrete methods, pending further study of clock rates for the
underlying gate operations and the ability of compilers to eliminate
redundant operations. As also observed with conventional heuristics,
the cost distribution for random $k$-SAT is highly skewed, so a few
instances dominate the mean cost. Instead, we use the median cost to
indicate typical behavior. The time for the continuous formulation is
$T = j \Delta$, so the adiabatic limit is $j \Delta \rightarrow
\infty$. By contrast, in the discrete formulation, $\Delta$
parameterizes the operators of \eq{step} rather than determining the
time required to perform them.

\eq{step} follows the continuous evolution, $\psi^{(h+1)} \approx
e^{-i H(f) \Delta} \psi^{(h)}$, when $\Delta ||H|| \rightarrow 0$
which holds when $\Delta \ll 1/n$~\cite{farhi00,vandam01}.  This last
condition uses the fact that the norm $||H||$ is the largest
eigenvalue of $H$, which is $\AtMost{n}$ since we consider $k$-SAT
problems with $m \propto n$.
As a specific choice, we use $\Delta = 1/\sqrt{j}$.
Other scaling choices $\Delta = 1/j^\alpha$ with $0<\alpha<1$ give
qualitatively similar behaviors to those reported here while
maintaining correspondence with the continuous evolution for
sufficiently large $j$.

The {\em unweighted} $\Hm$ uses equal weights: $w_i=1$ so
$\omega=n$. Alternatively, $w_i$ can be the number of times
variable $i$ appears in a clause~\cite{farhi01}, as also used by
some conventional heuristics to adjust the importance of changes
in each variable. This choice gives $\omega = m k$. By matching
$\Hm$ to the problem instance, one might expect such weights to
improve performance. Instead, for random 3-SAT these weights give
{\em higher} costs $C$, requiring about $n$ times as many steps to
achieve the same $\Psoln$ as the unweighted choice. If instead
these weights are normalized so their average value is 1, the
performance is about the same as in the unweighted case, but still
slightly worse. In light of these observations, we use the
unweighted $\Hm$ in this paper.

\begin{table}
\begin{center}
\begin{tabular}{l||c|c|l}
algorithm  & \multicolumn{2}{c|}{parameters} & phase functions \\
	& $T$ & $\Delta$ & \\ \hline \hline
adiabatic   & $T \rightarrow \infty$ & $\Delta \rightarrow 0$ & $\rho(0)=0=\tau(1)$  \\
discrete adiabatic & $T \rightarrow \infty$ & constant & $\rho(0)=0=\tau(1)$\\
heuristic   & constant & $\Delta \rightarrow 0$ & suitable $\rho$, $\tau$  \\
\end{tabular}
\end{center}
\caption{\label{algorithms} \small Summary of quantum search
algorithms using problem structure. The heuristic method requires
finding appropriate choices for the phase functions to give good
performance and for the number of steps $j$ to increase with problem
size $n$. The adiabatic methods require sufficiently large values of
$T=j \Delta$. A constant value for a parameter in this table means it
is taken to be independent of $n$ and $j$.}
\end{table}

We compare the adiabatic limit with two other methods, summarized in
Table~\ref{algorithms}. First, for the {\em discrete adiabatic} case
we take $\Delta$ independent of $n$ and $j$, violating the condition
$\Delta n \rightarrow 0$ so \eq{step} no longer closely approximates
the continuous evolution and does not necessarily give $\Psoln
\rightarrow 1$ as $j \rightarrow \infty$. In this case, a discrete
version of the adiabatic theorem, described in the appendix, ensures
$\Psoln$ is close to 1 if $\Delta$ is not too large.

Second, the {\em heuristic} method, studied
previously~\cite{hogg00,hogg01a}, has $\Delta=1/j$ and forms for
$\tau(f)$ and $\rho(f)$ that do not range between 0 and 1. Instead,
these phase functions must be selected appropriately to give good
performance. Identifying such choices and characterizing their
performance are major issues for this algorithm, though mean-field
approximations based on a few problem parameters, e.g., the ratio
$m/n$ for $k$-SAT, can give reasonably good choices. This method
does not correspond to the adiabatic limit: $\Psoln$ has a limit less
than 1 as $j \rightarrow \infty$.

For all these techniques, expected cost $C=j/\Psoln$ is minimized for
intermediate values of $j$ rather than taking $j \rightarrow \infty$
as used in the limits listed in Table~\ref{algorithms}. Identifying
parameters and phase functions, $\rho(f)$ and $\tau(f)$, giving
minimal cost for a given problem instance depends on details of the
search space structure unlikely to be available prior to solving
that instance. However, as described below, taking $j$ to grow only
as a fairly small power of $n$ provides relatively modest costs, on
average, for problem sizes feasible to simulate.

\section{BEHAVIOR}

\begin{figure}
\includegraphics[width=3in]{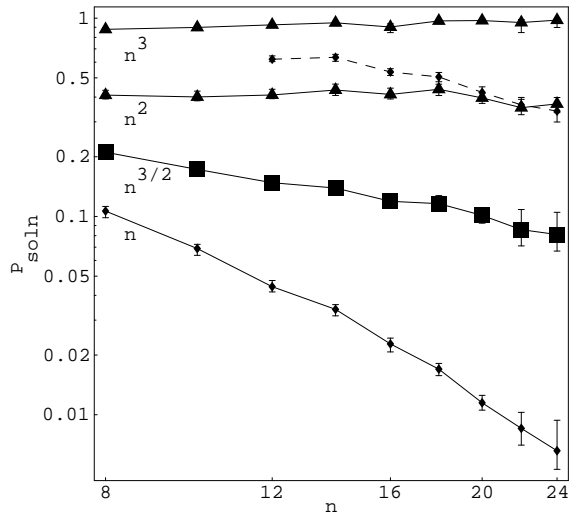}%
\caption[Probability of solution.]{\figlabel{pSoln} Log-log plot of
median $\Psoln$ for the adiabatic method vs.~$n$ with the number of
steps $j$ equal to $n$, the integer nearest $n^{3/2}$, $n^2$ and $n^3$
(solid curves, from bottom to top, respectively). We use
$\Delta=1/\sqrt{j}$. For comparison, the dashed curve shows $\Psoln$
for the heuristic method using at most $n$ steps. The error bars show
the 95\% confidence intervals~\cite[p.~124]{snedecor67} of the medians
estimated from the random sample of instances.
The same instances were solved with each method.  We use 1000
instances for each $n$ up to 20, and 500 for larger $n$, except only
100 for $j=n^3$ for $n\geq 16$.  }
\end{figure}

For the adiabatic method, \fig{pSoln} shows the median $\Psoln$ for
various growth rates of the number of steps. $\Psoln \rightarrow 1$ as
$j$ increases. At least for $n \lesssim 20$, $\Psoln \approx 1$ when
$j=n^3$,
so median costs are $\AtMost{n^3}$, a substantial improvement over all
known classical methods if it continues for larger $n$.  However, for
smaller powers of $n$, $\Psoln$ values decrease, but this is only
evident for $j=n^2$ for $n>20$. This raises the possibility of such a
decline, at somewhat larger $n$, for larger $j$ as well.  Provided
such a decline only leads to $\Psoln$ decreasing as a power of $n$,
corresponding to a straight line on the log-log plot of \fig{pSoln},
median costs would still only grow as a power of $n$. The remainder
of this section describes the algorithm behaviors in more detail.

\subsection{Energy Gap}

Asymptotically, the adiabatic method's cost is dominated by the growth
of $1/G^2$ where $G=\min_f g(f)$ and $g(f)$ is the energy gap in
$H(f)$, i.e., the difference between the ground state eigenvalue and
the smallest higher eigenvalue corresponding to a non-solution.
Evaluation using sparse matrix techniques~\cite{lehoucq98} for $n \leq
20$ gives the median $G$ in the range $0.3 - 0.5$, as illustrated for
one instance in \fig{gap}, and, more significantly, it does not
decrease over this range of $n$. This minimum is not much smaller than
other values of $g(f)$. Hence, unlike for large $n$, the cost is not
dominated by the minimum gap size and so the values of \fig{pSoln} may
not reflect asymptotic scaling.

\begin{figure}
\includegraphics[width=3in]{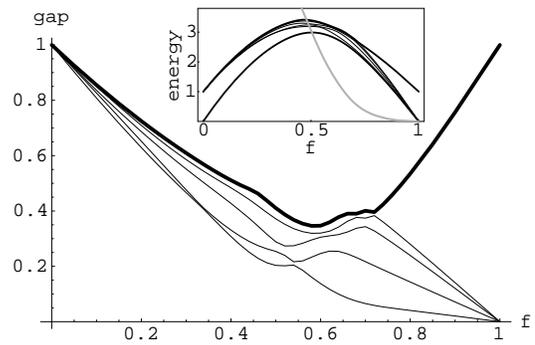}%
\caption[Energy gap.]{\figlabel{gap}
Difference between eigenvalues of the lowest 5 excited states and the
ground state vs.~$f$ for an instance with $n=20$, $m=85$ and 5
solutions. The inset shows the actual eigenvalues, with the gray curve
showing the expected cost $\expect{c}$ in the ground state.}
\end{figure}

\begin{figure}
\includegraphics[width=3in]{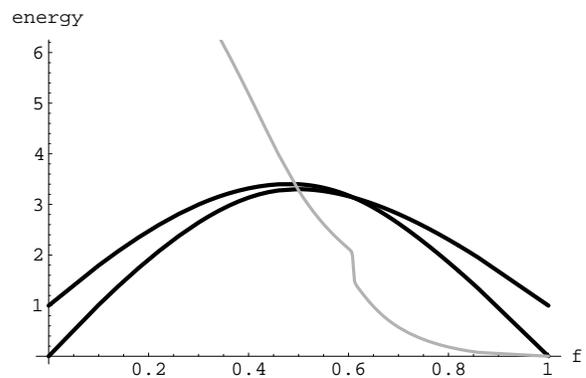}%
\caption[Average cost and gap.]{\figlabel{gap cost}
Lowest two eigenvalues vs.~$f$ for an instance with $n=20$, $m=85$,
one solution and a particularly small minimum gap. The gray curve
shows the expected cost $\expect{c}$ in the ground state, equal to
$m/2^k=10.625$ at $f=0$. Note the abrupt drop at the location of the
minimum gap. The ground state for $f=1$ is the solution, whose
cost is zero, so $\expect{c} \rightarrow 0$ as $f \rightarrow 1$.}
\end{figure}

By contrast, \fig{gap cost} illustrates the behavior of an instance
with a small minimum gap. One characterization of the eigenstates of
$H(f)$ is their expected cost, i.e., $\expect{c}_a = \sum_s c(s)
|\phi^{(a)}_s(f)|^2$ where $\phi^{(a)}(f)$ is the $a^{th}$ eigenvector
of $H(f)$. In particular, for $a=1$ this gives the expected cost in
the ground state, which we denote simply as $\expect{c}$.  The
expected cost in the ground state drops rapidly at the minimum gap
location, in contrast to the smooth behavior for instances with larger
gaps (as, for example, in \fig{gap}). We thus see a difference in
behavior of the ground state for instances with small gaps, presumably
representative of typical behavior for larger $n$, and the behavior of
more typical instances for $n \approx 20$.

With the adiabatic method and $T$ sufficiently large, the actual state
of the quantum computer after step $h$, $\psi^{(h)}$, closely
approximates the ground state eigenvector $\phi^{(1)}$, up to an
irrelevant overall phase. Thus the computation will also show the jump
in expected cost.

Detailed quantitative comparison of the typical behaviors due to small
minimum gaps and conventional heuristics requires larger problem
sizes.  Nevertheless, we can gain some insight from instances with
small gaps for $n \approx 20$, which tend to have high costs for both
the quantum methods and conventional heuristics, such as
GSAT~\cite{selman92}, even when restricting comparison to problems
with the same numbers of variables and solutions. For the instance
shown in \fig{gap cost}, GSAT trials readily reach states with 1 or 2
conflicts, but have a relatively low chance to find the solution. This
behavior, typical of conventional heuristics~\cite{frank97},
corresponds to the abrupt drop in $\expect{c}$ of \fig{gap cost}. Thus
finding assignments with costs below this value dominates the running
time of both the quantum and conventional methods. These observations
suggest small energy gaps characterize hard problems more generally
than just for the adiabatic method, which may provide useful insights
into the nature of search along with quantities such as the backbone
(i.e., variables with the same values in all
solutions~\cite{monasson99}).

Simple problems or algorithms ignoring problem structure allow
determining the gap for large
$n$~\cite{farhi00,vandam01,roland01}. This is difficult for random
SAT. For instance, although random $k$-SAT corresponds to random costs
for $\Hc$ and the extreme eigenvalues of random matrices can be
determined when elements are chosen
independently~\cite{edwards76,furedi81}, the costs of nearby states
for SAT instances are highly correlated since they likely conflict
with many of the same clauses.
Alternatively, upper~\cite{macdonald33} and lower~\cite{lowdin65}
bounds for eigenvalues can be based on classes of trial vectors. For
instance, vectors whose components for state $s$ depend only on $c(s)$
give fairly close upper bounds for the ground state of random 3-SAT,
on average, as well as a mean-field approximation for the
heuristic method~\cite{hogg00}. However, simple lower bounds for higher
energy states are below the upper bound for the ground state for some
values of $f$, and so do not give useful estimates for $G$.
Furthermore, typical soluble instances have exponentially many
solutions (although still an exponentially small fraction of all
states). Thus a full analysis of performance based on energy values
must also consider the behavior of the many eigenvalues corresponding
to solutions, which can be complicated, as illustrated in \fig{gap}.

\subsection{Search Cost}

\begin{figure}
\includegraphics[width=3in]{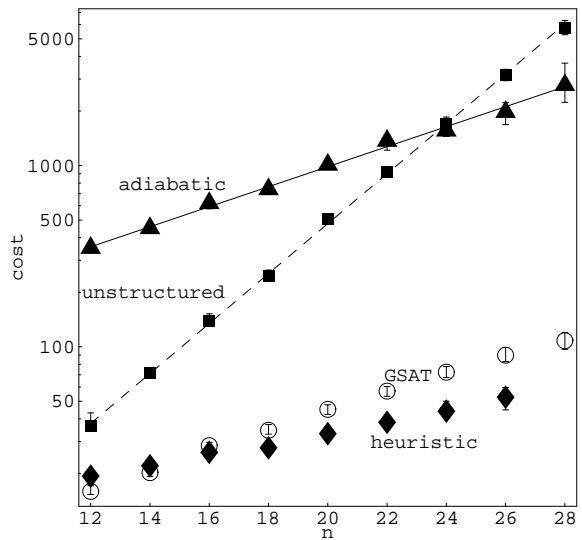}%
\caption[Median search costs.]{\figlabel{scaling} Log plot of median search cost vs.~$n$ for the heuristic (diamond), unstructured search (box), GSAT with restarts after $n$ steps (circle) and adiabatic search with $j=n^2$ (triangle).  The values are based on the same instances as in \fig{pSoln}. The lines are exponential fits to the unstructured (dashed) and adiabatic (solid) methods.}
\end{figure}
% GSAT with restarts after 2n steps has slightly larger costs but about the same rate of growth: exp(0.13 n)

Even if $n \lesssim 20$ does not identify asymptotic behavior, this
range of feasible simulations allows comparing algorithm costs. Such
comparisons are particularly relevant for quantum computer
implementations with relatively few qubits and limited coherence times
which are thus limited to small problems and few steps.  \fig{scaling}
compares the median values of the expected search costs $C$. For the
adiabatic method, using $j=n^3$ gives large costs, far higher than
those of conventional heuristics and other quantum methods. Using just
enough steps to achieve moderate values of $\Psoln$ reduces
cost~\cite{farhi01}, e.g., $j=n^2$. Alternatively, for each $n$,
testing various $j$ on a small sample of instances indicates the
number of steps required to achieve a fixed value of $\Psoln$, e.g.,
$1/8$. In our case, the latter approach has median costs about $20\%$
lower than the former, but with the same cost growth rate. Because this
improvement is minor compared to the differences with other algorithms
shown in the figure, and to avoid the additional variability due to
estimating $j$ from a sample of instances, we simply take $j=n^2$ to
illustrate the adiabatic method.

The figure also shows Grover's unstructured search~\cite{grover96}
(without prior knowledge of the number of solutions~\cite{boyer96})
and the conventional heuristic GSAT~\cite{selman92}.
\comment{Unlike the quantum methods, conventional heuristics can finish immediately when a solution is found rather than waiting until all $j$ steps are completed.
For comparison with the different choices of $j$ in \fig{pSoln}, the median costs at $n=20$ for $j=n$, $n^{3/2}$, $n^2$ and $n^3$ are, respectively, 1741, 879, 1010 and 8222.}
The unstructured search cost grows as $e^{0.32 n}$. The exponential
fit to the adiabatic method is $e^{0.13 n}$. This fit gives a
residual about half as large as that from a power-law fit.
The growth rate is about the same as that of GSAT.

\begin{figure}
\includegraphics[width=3in]{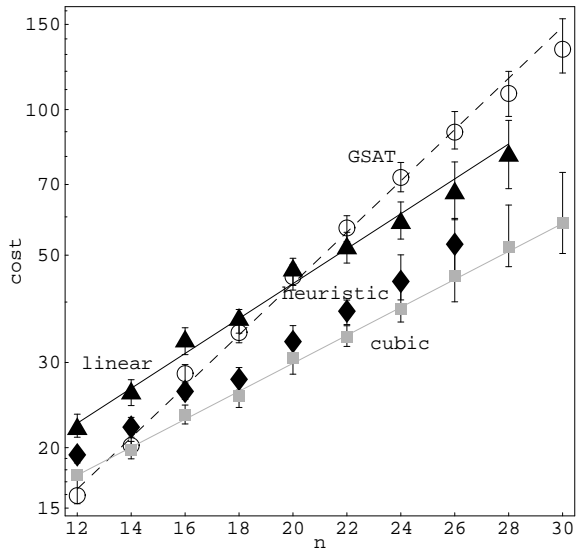}%
\caption[Median search costs.]{\figlabel{scaling1} Log plot of median
search cost vs.~$n$ for GSAT (circle), the heuristic method (diamond),
both of which are also shown in \fig{scaling}, and two versions of the
discrete adiabatic method: $\Delta=1.2$ with linear phase functions
(triangle) and the cubic polynomial variation with $f$ (gray box)
described in the text.  The lines are exponential fits to GSAT
(dashed) and the two discrete adiabatic quantum methods. The figure
uses the same instances as \fig{scaling}.}
\end{figure}

\fig{scaling} shows the heuristic, using at most $n$ steps, gives low
costs due to its fairly high values for $\Psoln$ shown in \fig{pSoln}.
The constant $\Delta$ scaling for the discrete adiabatic method also
gives large $\Psoln$ values for $j \approx n$. Thus both $\Delta =
1/j$ and $\Delta$ independent of $j$ make better use of quantum
coherence in the discrete formulation than the continuous adiabatic
limit (with $1/j \ll \Delta \ll 1/n$) for hard random 3-SAT. These
behaviors are shown in \fig{scaling1}.

Because these quantum methods and GSAT consist of a series of
independent trials, they can be combined with amplitude amplification
to give an additional quadratic performance
improvement~\cite{brassard98}. However, this is only a significant
benefit when $\Psoln$ is fairly small, which is not the case for the
heuristic and GSAT methods for these problem sizes.

For the adiabatic method, taking $\rho(f)$ and $\tau(f)$ in \eq{step} to
vary according to $g(f)^2$ reduces costs~\cite{vandam01,roland01}.
This concentrates steps at values of $f$ close to the minimum gap.
While $g(f)$ is costly to evaluate for SAT instances,
using average values of $g(f)$ based on a sample
of instances gives some benefit. E.g., for $j=n^2$, $\Psoln$
increases from around $0.4$ shown in \fig{pSoln} to a range of $0.5 -
0.6$ but this does not appear to reduce the cost's growth rate.

Similar improvement occurs with constant $\Delta$.
Optimizing $\tau$ and $\rho$ separately for each step on a sample
of instances gives values close to a cubic polynomial in $f$.
Restricting attention to such polynomials, for a set of 100 $n=12$
instances the best performance was with $\Delta=1.31275$,
$\rho(f)=p(f)$, $\tau(f)=1-p(f)$ where $p(f)=1.92708 f - 2.66179 f^2 +
1.73471 f^3$.
This cubic is similar to the functional form optimizing the
adiabatic method for unstructured search~\cite{roland01,vandam01}.
\fig{scaling1} shows the resulting cost reduction. Hence, tuning the
algorithm to the problem ensemble is beneficial, as also suggested
by a mean-field analysis of the heuristic~\cite{hogg00}.

The simulations also show these quantum algorithms have a large
performance variance among instances with given $n$ and $m$, and no
single choice for $\rho$ and $\tau$ is best for all problem instances. Thus
portfolios~\cite{maurer01} combining a variety of such
choices can give further improvements.

\section{CONCLUSION}

In summary, for random SAT, the adiabatic method improves on
unstructured search and provides a general technique to exploit
readily computed properties of hard search problems through the choice
of Hamiltonians. However nonadiabatic-limit algorithms require fewer
steps, comparable to GSAT, and appear to have slower cost growth. As a
caveat, small energy gaps appear to be associated with instances
difficult to solve with both quantum and classical methods. Thus the
simulation results presented here, based on fairly small problem sizes
for which most instances have fairly large energy gaps, may not reveal
the asymptotic scaling of the typical search cost for hard random SAT
problems. Evaluating the behavior of these algorithms and, more
generally, identifying better ways to use state costs in quantum
algorithms remain open questions.

Quantum computers with only a moderate number of qubits could test
algorithms beyond the range of simulators, and hence provide useful
insights even if the problem sizes are still readily solved by
conventional heuristics.  Such studies could help address the question
of whether, with suitable tuning based on readily evaluated average
properties of search states, the ability to operate on the entire
search space allows quantum computers to effectively exploit weak
correlations among state costs in ways classical machines cannot.

\begin{acknowledgments}
I have benefited from discussions with Rob Schreiber and Wim van Dam. I thank Miles Deegan and the HP High Performance Computing Expertise Center for providing computational resources for the simulations.
\end{acknowledgments}

\appendix

\section{DISCRETE ADIABATIC BEHAVIOR}
{
\newcommand{\ev}{ {\hat{e}} }
\newcommand{\et}{ {\hat{e}^\dagger} }

When $\Delta$ is held constant, the steps of \eq{step} do not
approximate the continuous evolution induced by $H(f)$, and hence
$\psi^{(h)}$ does not closely follow the ground state of $H(f)$ when
$T \rightarrow \infty$. Nevertheless, $\psi^{(h)}$ does closely follow
an eigenstate of the unitary operator involved in \eq{step}. This
discrete version of the adiabatic theorem ensures good performance of
the algorithm provided the continuous change in the eigenvector takes
the initial ground state into the final one, rather than into some
other eigenvector.

\subsection{The Discrete Adiabatic Limit}

Consider a smoothly changing sequence of unitary matrices $U(f)$ defined
for $0 \leq f \leq 1$ and vectors $\psi^{(h+1)} = U(f) \psi^{(h)}$
with $f = h/j$ for $h=0,\ldots,j-1$. Let $e^{-i \theta_r(f)}$ and $\ev_r(f)$
be the $r^{th}$ eigenvalue and (normalized) eigenvector of $U(f)$. 

We start with $\psi^{(0)}$ equal to the eigenvector $\ev_1(0)$ of
$U(0)$, which we assume to be nondegenerate for simplicity. Provided
the difference between eigenvalues is bounded away from zero, for
sufficiently large $j$, $\psi^{(j)}$ will be close to an eigenvector
of $U(1)$. To see this let $\epsilon = 1/j$ and expand $\psi^{(h)} =
\sum_r c_r(f) \Lambda_r(f) \; \ev_r(f)$ in the eigenbasis of $U(f)$
where
\begin{displaymath}
\Lambda_r(h/j) \equiv  \exp \left( -i \sum_{k=0}^{h-1} \theta_r(k/j) \right)
\end{displaymath}

First order perturbation theory gives the change in the $c_r$ values
during one step to be $\AtMost{\epsilon}$. After $j$ steps, it might
appear that these changes could build up to $\AtMost{\epsilon
j}=\AtMost{1}$. However, this is not the case due to the rapid
variation in phases when $j$ is large. Specifically, the changes
in coefficients for $r \neq 1$ are
\begin{equation}\eqlabel{diff eqn}
\frac{d c_r}{d f}  = P_{1,r}(f) \Phi_r(f)
\end{equation}
where $P_{s,r}(h/j) \equiv e^{-i j \;\Theta_{s,r}(f)}$,
\begin{displaymath}
\Theta_{s,r}(f) \equiv
    \frac{1}{j} \sum_{k=0}^{h-1} ( \theta_s(k/j) - \theta_r(k/j) )
\end{displaymath}
and 
\begin{displaymath}
\Phi_r \equiv \frac{\langle r| \, dU/df \, |1 \rangle}{e^{-i \theta_r}-e^{-i \theta_1}}
\end{displaymath}

Since $c_r(0)=0$, \eq{diff eqn} gives
\begin{displaymath}
c_r(f) = \int_0^f e^{-i j \, \Theta_{1,r}(\kappa)} \Phi_r(\kappa) \;
d \kappa
\end{displaymath}
As $j$ increases, the integrand oscillates increasingly rapidly so the
integral goes to zero as $j \rightarrow \infty$ by applying the
Riemann-Lebesgue lemma since $d \Theta_{1,r}/d f = \theta_1 -
\theta_r$ is nonzero and $|\Phi_r(f)|$ is bounded for all $f$ and $r
\neq 1$, by the assumption of no level crossing.  Hence $c_r(f)
\rightarrow 0$ so $\psi^{(j)}$ approaches $\ev_1(1)$, up to an overall
phase factor, as $j \rightarrow \infty$.

\subsection{An Example}

\begin{figure}[!t]
\includegraphics[width=3in]{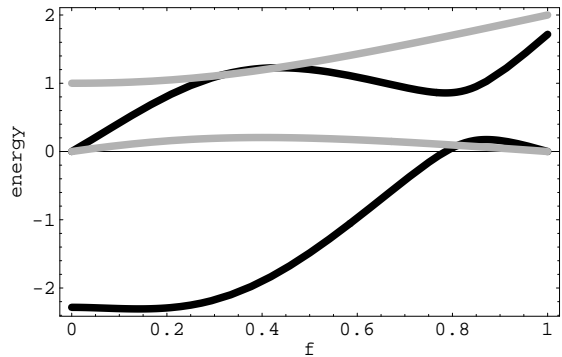}%
\caption[Example of eigenvalue behavior.]{\figlabel{evExample} Energy
values $\theta_r(f)$ corresponding to the two eigenvalues of $U(f)$
vs.~$f$ for $\Delta=1$ (gray) and 4 (black). The values are defined
only up to a multiple of $2\pi$, and we take $-\pi < \theta \leq
\pi$. The ground states of $\Hm$ and $\Hc$ correspond to $\theta(0)=0$
and $\theta(1)=0$, respectively. The values for $\Delta=1$ are close
to those of the combined Hamiltonian $H(f)=(1-f)\Hm + f \Hc$. However,
the $\Delta=4$ values do not remain close to those of $H(f) \Delta$.}
\end{figure}

\begin{figure}[!th]
\includegraphics[width=3in]{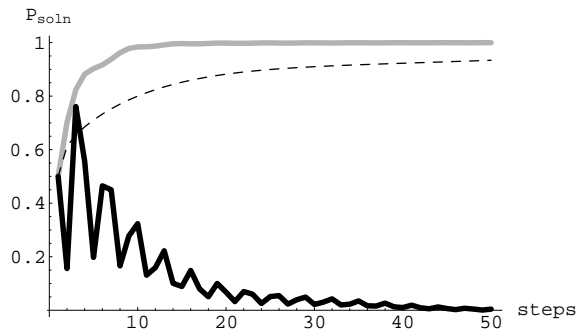}%
\caption[Example of discrete adiabatic method.]{\figlabel{pSolnExample} 
$\Psoln$ vs.~$j$ for $\Delta=1$ (gray) and 4 (black). For comparison, the dashed curve uses $\Delta=1/\sqrt{j}$ corresponding to the continuous adiabatic limit.}
\end{figure}

An important caveat in applying this result to quantum algorithms is
that while $j \rightarrow \infty$ suffices to ensure $\psi^{(h)}$
closely follows the evolution of an eigenvector of $U(f)$, this
evolution may not lead to the desired eigenvector of $U(1)$, i.e.,
corresponding to solutions to the search problem. This is because the
eigenvalues of $U(f)$ lie on the unit circle in the complex plane and
can ``wrap around'' as $\Delta$ increases. Hence, in addition to
ensuring the eigenvalue gap does not get too small, good performance
also requires selecting appropriate $\Delta$. Alternatively, one could
start from a different eigenvector of $U(0)$, which would be useful
if one could determine which eigenvector maps to the solutions.

One guarantee of avoiding this problem is that {\em none} of the
eigenvalues of $U(f)$ wrap around the unit circle, i.e., $\Delta ||H||
\rightarrow 0$, corresponding to the continuous adiabatic
limit. Simulations show performance remains good for moderate values
of $j$ even if $\Delta$ does not go to zero, provided $\Delta$ is
below some threshold value. For hard random 3-SAT problems with $j
\propto n$, this threshold appears to be somewhat larger than 1.

To illustrate these remarks consider the $n=1$ example 
\begin{displaymath}
\Hm = \frac{1}{2} \pmatrix{1 & -1 \cr -1 & 1}, \;\hfill\;
\Hc = \pmatrix{0 & 0 \cr 0 & 2}
\end{displaymath}
so $U(f) = e^{-i \Hm (1-f) \Delta} e^{-i \Hc f
\Delta}$. \fig{evExample} shows the behavior of the two eigenvalues of
$U(f)$ for two values of $\Delta$. For $\Delta=4$ the initial ground
state eigenvector, with eigenvalue 1, evolves into the $2^{nd}$
eigenvector of $U(1)$ rather than the eigenvector corresponding to the
ground state of $\Hc$.

\fig{pSolnExample} shows the consequence of this behavior: when
$\Delta$ is too large, $\psi^{(h)}$ follows the evolving eigenvector
to the wrong state when $f=1$, giving $\Psoln \rightarrow 0$ as $j
\rightarrow \infty$. As another observation from this figure,
$\Psoln(j)$ exhibits oscillations (though they are quite small for
$\Delta=1$). With appropriate phase choices, these oscillations can be
quite large, allowing $\Psoln$ to approach 1 with only a modest number
of steps, even when $\Psoln$ approaches 0 for larger $j$. This
observation is the basis of the heuristic method.

\begin{figure}[t]
\includegraphics[width=3in]{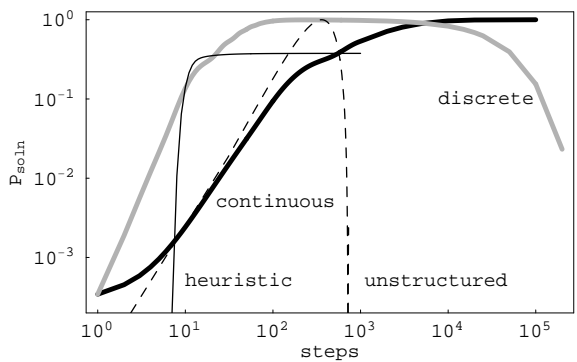}%
\caption[Comparing methods for a 3-SAT instance.]{\figlabel{pSolnCompare} 
$\Psoln$ vs.~$j$ for several search methods solving a 20-variable 3-SAT instance with 85 clauses and 5 solutions, the same instance used in \fig{gap}. The gray curve is the discrete adiabatic method with $\Delta=1$, the thick black curve is for $\Delta=1/\sqrt{j}$ corresponding to the continuous adiabatic limit. For comparison, the thin black curve is the heuristic, with $\Delta=1/j$, and the dashed curve is for unstructured search (showing only the first period of its sinusoidal oscillation on this log-log plot).}
\end{figure}

To see the consequence of this behavior for search, \fig{pSolnCompare} compares the behavior of several search methods. In this case $\Delta=1$ is sufficiently large that the initial eigenstate of the unitary operator evolves into a nonsolution eigenstate. Thus as the number of steps increases, the probability to find a solution goes to zero, as with the large $\Delta$ case in \fig{pSolnExample}. Nevertheless, for smaller $j$, most of the amplitude ``crosses'' the gap to another eigenstate that does evolve to a solution state. Consequently, this discrete adiabatic method gives lower overall search cost, using a moderate number of steps, than the continuous adiabatic method (which has $\Psoln \rightarrow 1$ as $j \rightarrow \infty$). By contrast, for $\Delta$ larger than 2 or so $\Psoln$ always remains small.

} % end of group for appendix

\end{document}